\documentclass[letter]{aa}
\usepackage{graphicx}
\usepackage{amsmath}
\usepackage{txfonts}
\usepackage{subfigure}

\usepackage{natbib}
\bibpunct{(}{)}{;}{a}{}{,} 

\usepackage{hyperref}

\graphicspath{{figures/}}

 \title{Turbulent transport in radiative zones of stars}
\author{V. Prat \and F. Ligni\`eres}
\institute{CNRS; IRAP; 14, avenue \'Edouard Belin, F-31400 Toulouse, France \and Universit\'e de Toulouse; UPS-OMP; IRAP; Toulouse, France}
\offprints{V. Prat}
\mail{vprat@irap.omp.eu}
\date{Received 18 October 2012 / Accepted 17 January 2013}
\abstract
{In stellar interiors, rotation is able to drive turbulent motions, and the related transport processes have a significant influence on the evolution of stars.
Turbulent mixing in the radiative zones is currently taken into account in stellar evolution models through a set of diffusion coefficients that are generally poorly constrained.}
{We want to constrain the form of one of them, the radial diffusion coefficient of chemical elements due to the turbulence driven by radial differential rotation, derived by Zahn (1974, 1992) on phenomenological grounds and largely used since.}
{We performed local, direct numerical simulations of stably stratified homogeneous sheared turbulence using the Boussinesq approximation.
The domain of low P\'eclet numbers found in stellar interiors is currently inaccessible to numerical simulations without approximation.
It is explored here thanks to a suitable asymptotic form of the Boussinesq equations.
The turbulent transport of a passive scalar is determined in statistical steady states.}
{We provide a first quantitative determination of the turbulent diffusion coefficient and find that the form proposed by Zahn is in good agreement with the results of the numerical simulations.}
{}

\begin{document}

\maketitle

\section{Introduction}

As many astrophysical measures are calibrated on stellar evolution theory, realistic stellar evolution models are crucial for astrophysics.
One of the major issues is the influence of the macroscopic motions induced by the rotation of a star on its internal structure \citep[see the recent review of][]{Maeder2012}.
In particular, chemical mixing can significantly extend the lifetime of the stars by continuously providing fresh combustible matter to the core, thus feeding nuclear reactions for a longer time.

A key point of these stellar evolution models is the radial turbulent mixing of chemical species induced by radial differential rotation.
According to Zahn's model \citep{Zahn1974}, this turbulent mixing is described by a diffusion coefficient that reads as
\begin{equation} \label{eq_D_Zahn}
 D_{\rm t}=\frac{Ri_{\rm cr}}{3}\kappa\left(\frac{r\sin\theta}{N_{\rm T}}\frac{{\rm d}\Omega}{{\rm d}r}\right)^2,
\end{equation}
where $\kappa$ is the thermal diffusivity, $Ri_{\rm cr}$ the critical value of the Richardson number, $N_{\rm T}$ the Brunt-Väisälä frequency, and $r\sin\theta{\rm d}\Omega/{\rm d}r$ the shear associated with radial differential rotation.
This expression is taken from Eq.~(2.14) of \cite{Zahn1992}.
While other versions of this diffusion coefficient exist in the literature, notably those accounting for additional effects like the $\mu$-gradient \citep{Maeder1996,Talon,Mathis}, Zahn's prescription is the basis of most implementations of the turbulent mixing driven by differential rotation in stellar evolution codes \citep[see a detailed discussion in \citealp{Meynet2000}]{Endal, Pinsonneault, Heger}.
Moreover, this mixing plays a significant role as compared to other transport processes, such as the meridional circulation combined with horizontal turbulence, especially for massive stars \citep{Meynet1997, Meynet2000, Denissenkov} and low-mass stars with strong $\mu$-gradients \citep{Palacios2003, Palacios2006}.
It thus has a direct impact on the confrontations of stellar evolution models with observations.
While chemical abundances at the star surface have been widely used for such confrontations \citep{Michaud, Brott, Potter}, asteroseismology is now providing direct constraints on these processes, especially for subgiant and giant stars \citep{Deheuvels, Eggenberger}.

The derivation of Zahn's prescription is based on the hypothesis that turbulent flows are statistically stationary and that this is achieved when the mean shear flow is close to the linear marginal stability criterion.
Following \cite{Townsend}, Zahn modified Richardson's instability criterion to take the destabilizing effect of thermal diffusivity into account in such a way that marginal stability is controlled by $RiPe \sim 1$, where the P\'eclet number $Pe$ compares the diffusive time scale with the dynamical one (as later confirmed through numerical linear stability analyses, \emph{e.g.} \citealt{Lignieres1999b}). 
Then, he assumed that, in a turbulent state, the relevant P\'eclet number in the expression $RiPe \sim 1$ is the turbulent one $Pe_\ell=\frac{u\ell}{\kappa}$, defined with the turbulent scales of velocity and length ($u$ and $\ell$, respectively). The diffusion coefficient is then estimated through the relation $D=u\ell/3$, which finally leads to Eq.~\eqref{eq_D_Zahn}.

The purpose of our present work is to use local 3D direct numerical simulations (DNS) to test Zahn's modeling of turbulent mixing.
Previous numerical simulations performed in a geophysical context have shown that a steady stably stratified homogeneous sheared turbulence can be obtained by imposing both a uniform mean shear and a uniform stratification and by tuning the Richardson number to reach statistical stationarity \citep{Jacobitz}.
Such a flow is well adapted to studying chemical turbulent transport and enables one to relate the transport to the local gradients of velocity and entropy as in Zahn's criterion.
Imposing the shear and the stratification amounts to an assumption of scale separation between the variations in the mean and turbulent flows. 
The main difficulty of performing such numerical simulations in a stellar context is the very low Prandtl number of the stellar fluid.
Indeed, this introduces a huge gap between the dynamical time scale and the diffusive one, which forces use of a large number of numerical time steps to accurately compute the effect of thermal diffusion on a dynamical time.
This would require a prohibitive amount of computation time for DNS, where the whole range of scales of turbulence down to the dissipative scales are resolved.
An attempt to test Zahn's prescription was made by \cite{Bruggen}, but in this case the simulations relied on the numerical diffusivity of the ZEUS code that is strongly dependent on the grid size \citep{Fromang}.
Fortunately, an asymptotic form of the Boussinesq equations in the limit of the small P\'eclet numbers exists, where the numerical constraint associated with the dynamical effect of the high thermal diffusivity disappears \citep{Lignieres1999a}.

In this paper, we perform DNS of 3D steady, stably stratified, homogeneous sheared turbulence at decreasing P\'eclet numbers and use both the Boussinesq equations and the so-called small-P\'eclet-number approximation (SPNA) to explore the asymptotic $Pe\ll1$ regime.
The mathematical model of the considered turbulent flow is presented in Sect.~\ref{sec_flow}, followed by the numerical simulations (Sect.~\ref{sec_simulation}) and the methods used to study the turbulent transport (Sect.~\ref{sec_transport}).
Then, in Sect.~\ref{sec_results}, constraints on the diffusion coefficient obtained from our numerical simulations are described and compared to Zahn's prescription.
The results are discussed in Sect.~\ref{sec_conclusion}.

\section{Mathematical model of the flow} \label{sec_flow}

\begin{figure}
 \resizebox{\hsize}{!}{\includegraphics{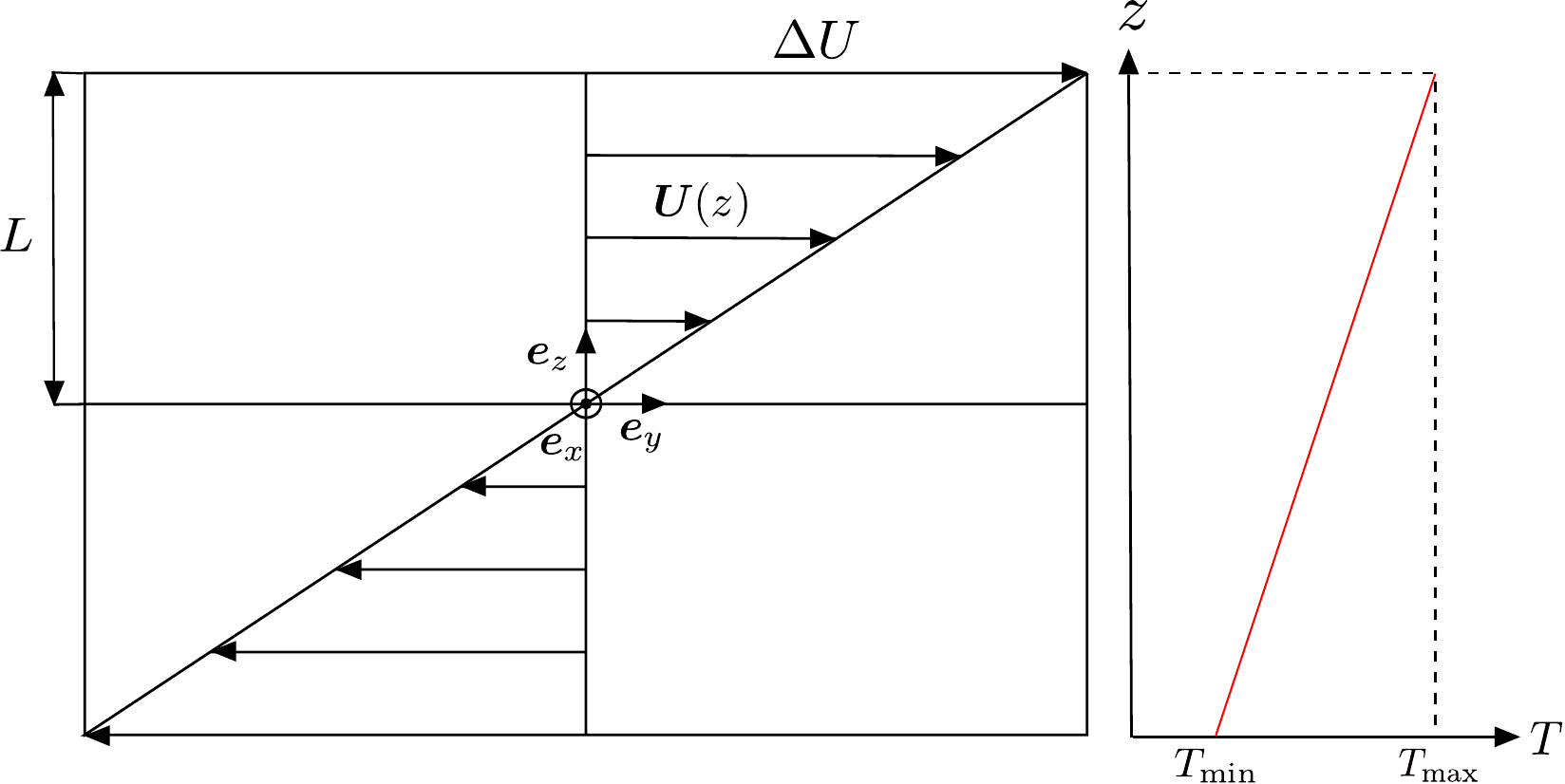}}
 \caption{Sketch of the flow configuration.}
 \label{fig_flow}
\end{figure}

To obtain results that are as generic as possible, we consider the simplest flow likely to produce stably stratified, homogeneous, sheared turbulence. Its characteristics are a uniform vertical shear of horizontal velocity and a uniform vertical temperature gradient, as shown in Fig.~\ref{fig_flow}.
The governing equations of the flow are presented in Sect.~\ref{sec_eq}, and the boundary conditions along with the forcing and the initial conditions in Sect.~\ref{sec_bound}.

\subsection{Governing equations} \label{sec_eq}

We use the Boussinesq equations in which density fluctuations are neglected, except in the buoyancy term.
For the usually low Mach-number flows inside stars, this approximation is justified as long as the motion vertical length scale is smaller than the pressure scale height.
Using $L$, $\Delta U$, and $\Delta T=(T_{\rm max}-T_{\rm min})/2$ as length, velocity, and temperature units (see Fig.~\ref{fig_flow}), their non-dimensional form reads as
\begin{eqnarray}
\vec\nabla\cdot\vec{\tilde u}									&=&	0, \\
\frac{\partial\vec{\tilde u}}{\partial\tilde t}+\vec{\tilde u}\cdot\vec\nabla\vec{\tilde u}		&=&	-\vec\nabla\tilde p+Ri\tilde\theta\vec e_z+\frac{1}{Re}\Delta\vec{\tilde u}+\vec{\tilde f}_{\rm v},	\label{eq_momentum}\\
\frac{\partial\tilde\theta}{\partial\tilde t}+\vec{\tilde u}\cdot\vec\nabla\tilde\theta+\tilde w	&=&	\frac{1}{Pe}\Delta\tilde\theta+\tilde f_{\rm T},	\label{eq_temperature}
\end{eqnarray}
where $\vec{\tilde u}=\tilde u\vec e_x+\tilde v\vec e_y+\tilde w\vec e_z$ is the velocity, $\tilde p$ the pressure deviation from hydrostatic equilibrium, $\tilde\theta$ the temperature deviation from the mean profile, and $\vec{\tilde f}_{\rm v}$ and $\tilde f_{\rm T}$ forcing terms (see Sect.~\ref{sec_bound}).

These equations show three non-dimensional parameters: (i)~the Richardson number $Ri=(N_{\rm T}/S)^2$, noting $S={\rm d}U/{\rm d}z=\Delta U/L$ the shear rate, ${N_{\rm T}}^2=\alpha g\Delta T/L$ the Brunt-Väisälä frequency, and $\alpha$ the thermal expansion coefficient, (ii)~the ``macroscopic'' Reynolds number $Re=L\Delta U/\nu=SL^2/\nu$, where $\nu$ is the molecular viscosity, and (iii)~ the ``macroscopic'' P\'eclet number $Pe=L\Delta U/\kappa=SL^2/\kappa$.

To explore the regime of very low P\'eclet numbers, we use the SPNA \citep{Lignieres1999a}.
The basic principle of this approximation is to Taylor-expand all variables up to first order in $Pe$.
Thus, Eqs.~\eqref{eq_momentum} and \eqref{eq_temperature} become
\begin{eqnarray}
\frac{\partial\vec{\tilde u}}{\partial\tilde t}+\vec{\tilde u}\cdot\vec\nabla\vec{\tilde u}	&=&	-\vec\nabla\tilde p+RiPe\tilde\psi\vec e_z+\frac{1}{Re}\Delta\vec{\tilde u}+\vec{\tilde f}_{\rm v}, \\
\tilde w										&=&	\Delta\tilde\psi,
\end{eqnarray}
with $\tilde\psi=\tilde\theta/Pe$.
The background thermal stratification remains fixed because temperature deviations become infinitely small as $Pe$ approaches zero.
There are two non-dimensional parameters left: the Reynolds number and the ``Richardson-P\'eclet'' number $RiPe$, which compares the effect of the stratification modified by the thermal diffusion (with a time scale of ${\tau_{\rm B}}^2/\tau_\kappa$, where $\tau_{\rm B}=1/N_{\rm T}$ and $\tau_\kappa=L^2/\kappa$ are the buoyancy and diffusive time scales, respectively) to that of the shear.
This number is known to control the linear stability of shear flows at low P\'eclet number \citep[][and references herein]{Lignieres1999b}.

An expansion similar to the SPNA has been proposed for an unstable thermal stratification \citep{Thual}.
However, it does not allow the feedback of the convective turbulent motions on the stratification, as observed in the solar convective zone.

The numerical code uses a Fourier colocation method in the horizontal directions, compact finite differences in the vertical, and a projection method plus Runge-Kutta for time advancing.

\subsection{Boundary conditions, forcing, and initial conditions} \label{sec_bound}

Periodic boundary conditions apply in the horizontal directions, whereas in the vertical direction the fluid is bounded by two horizontal surfaces that are impenetrable ($w=0$), stress-free in the spanwise direction ($\partial_z u = 0$), uniformly sheared in the streamwise direction ($\partial_z v = S$), and thermally controlled ($\theta=0$).
The mean shear profile $\vec U(z)$ is imposed thanks to a body force $\vec f_{\rm v} = [\vec U(z) - \bar v\vec e_y]/\Delta t$ applied at every time step $\Delta t$, where $\bar v$ is the horizontal average of the streamwise velocity.
The mean vertical temperature profile is imposed in the same way.
With such a set-up, turbulence is not rigorously homogeneous in the vertical direction.
Nevertheless, as already found in \cite{Schumacher} using similar forcing and boundary conditions, non-homegeneity is limited to thin layers ($\sim10\%$ of the domain vertical extent) close to the upper and lower boundaries.

As initial conditions, we construct a statistically homogeneous and isotropic Gaussian random velocity fluctuation field \citep{Orszag} with an energy spectrum proportional to $k^2e^{-k^2/{k_0}^2}$, where $k$ is the norm of the wave vector and $k_0$ corresponds to the maximum of the spectrum \citep{Jacobitz}.
The initial temperature deviation $\theta$ is set to zero.

\section{Numerical simulations} \label{sec_simulation}

\begin{figure}
 \resizebox{\hsize}{!}{\includegraphics{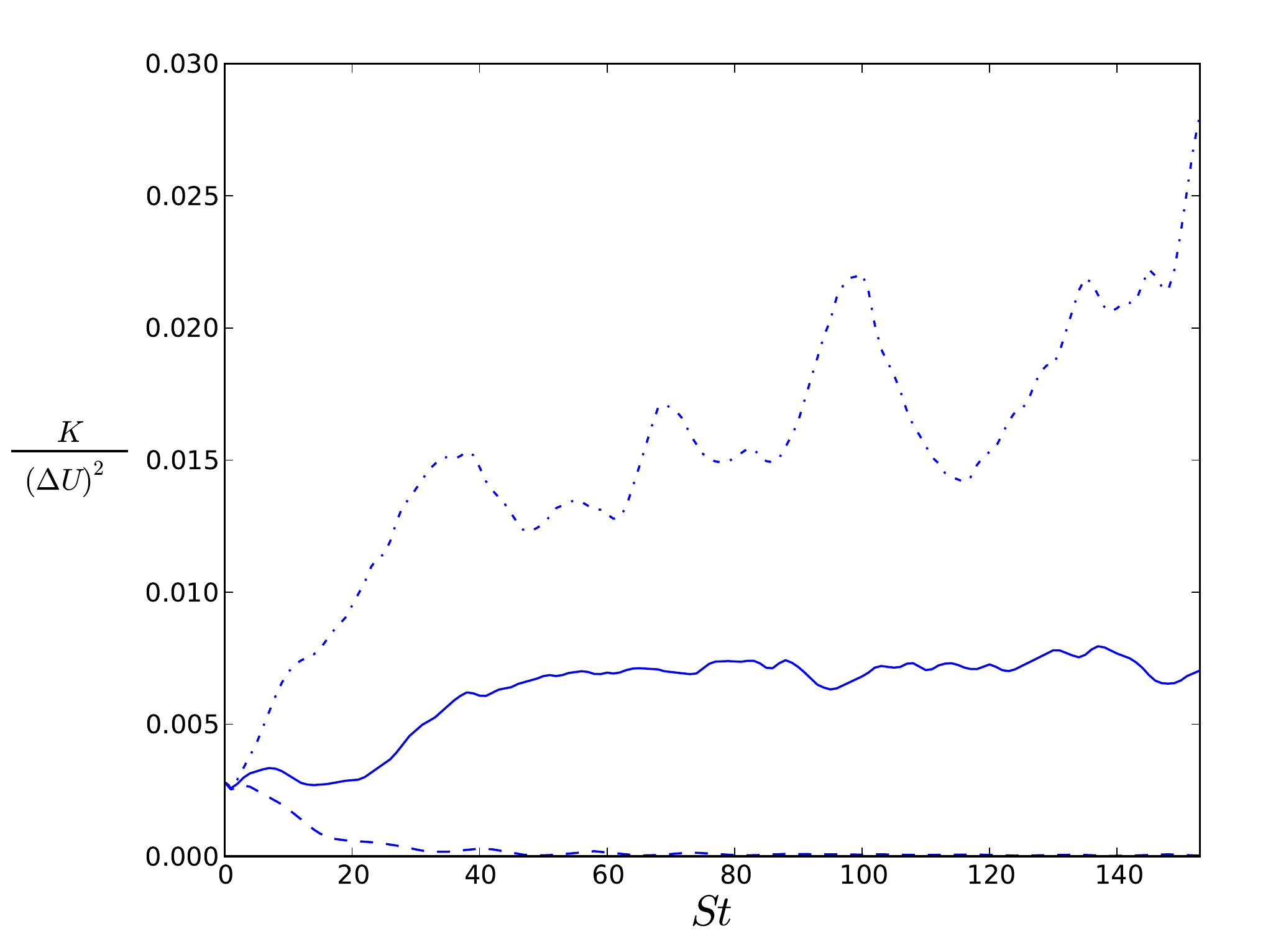}}
 \caption{Evolution of the turbulent kinetic energy $K$ for three different Richardson numbers at turbulent P\'eclet number $Pe_\ell=52$.
After a transient regime, the turbulent kinetic energy is stationary for $Ri=0.124$, increases for $Ri=0.05$ and decreases for $Ri=0.2$.
 }
 \label{fig_evolution}
\end{figure}

For each value of the P\'eclet number, the Richardson number can be tuned to obtain statistical steady-state flows.
In the regime where thermal diffusivity has no dynamical effect, various studies  \citep{Holt,Jacobitz} had already shown there is 
a critical Richardson number such that the turbulent kinetic energy increases if $Ri<Ri_{\rm cr}$ and decreases if $Ri>Ri_{\rm cr}$, as illustrated in Fig.~\ref{fig_evolution}.
We could also find such critical Richardson numbers in the low-P\'eclet-number regime.

\begin{figure}
 \resizebox{\hsize}{!}{
 \subfigure{\includegraphics{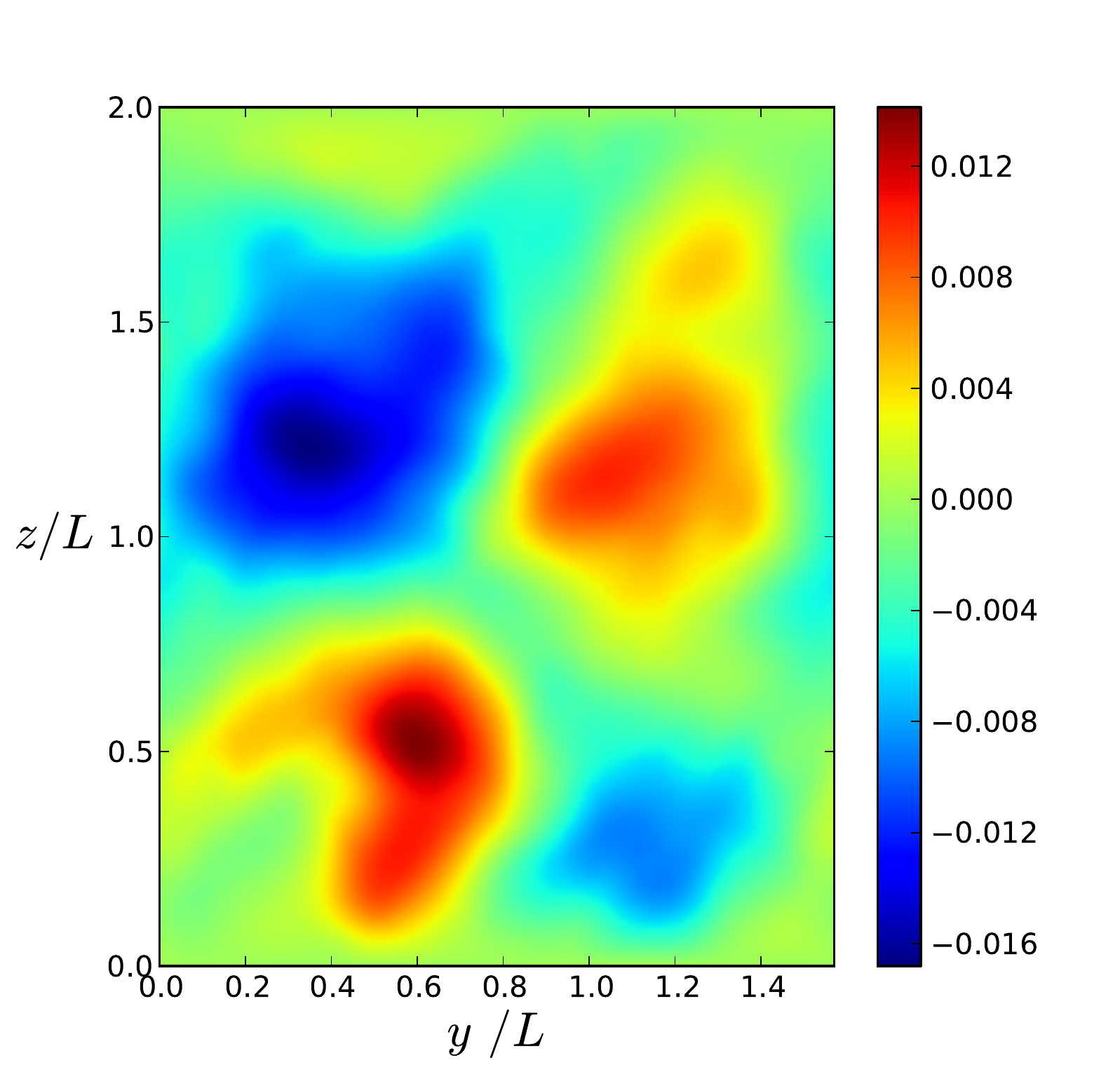}}
 \subfigure{\includegraphics{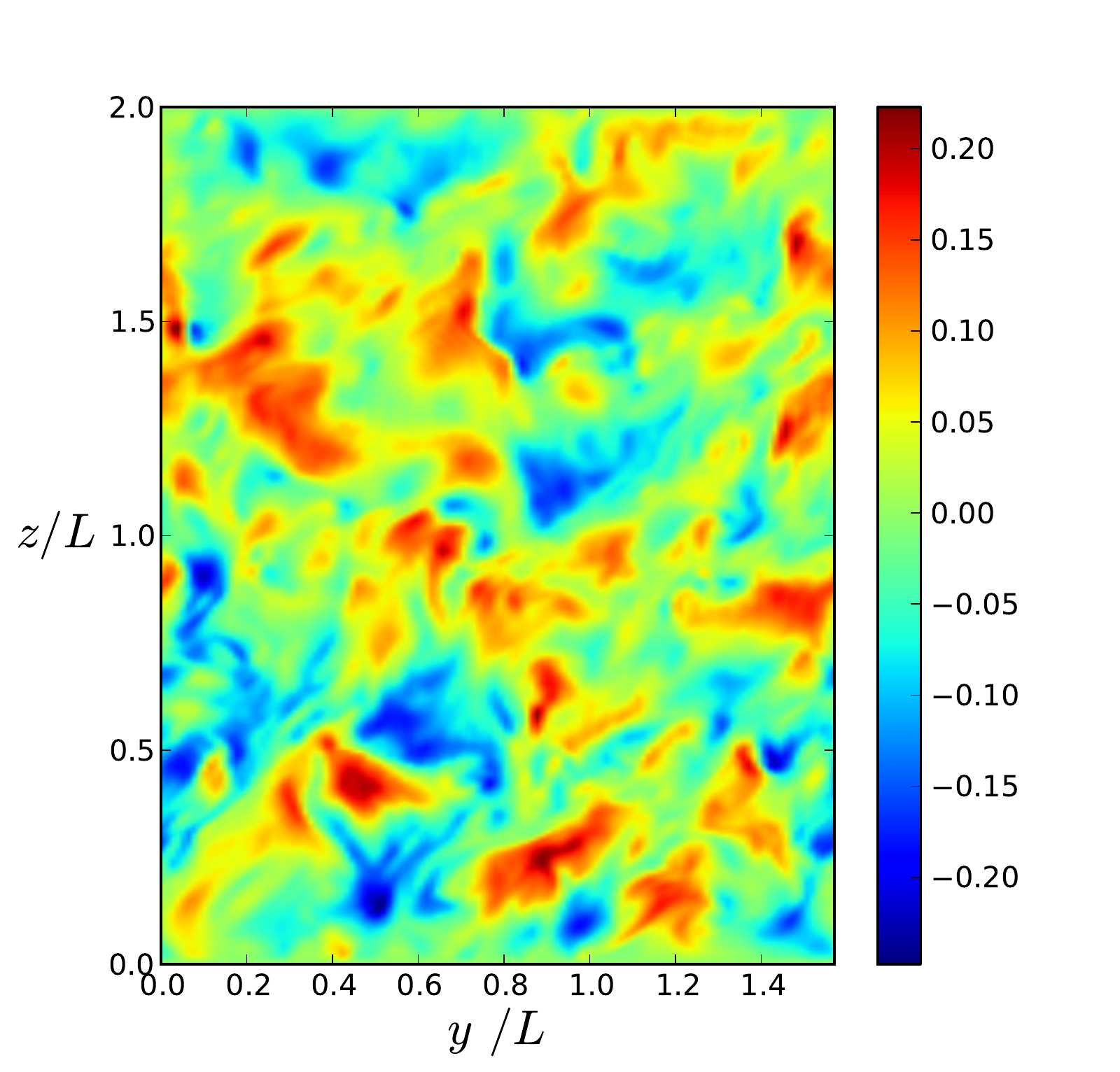}}
 }
 \caption{
    Snapshots of temperature fluctuations scaled by $\Delta T$ (left) and vertical velocity scaled by $\Delta U$ (right) in the $yOz$ plane at $Pe_\ell=0.34$.
    Vertical velocity presents smaller scales than temperature fluctuations; both are anti-correlated.
 }
 \label{fig_temperature}
\end{figure}

Moreover, to obtain reliable statistics of the turbulent transport, the simulation domain must contain several correlation length scales of the turbulence and the duration of the simulation must be long compared to its correlation time scale.
Denoting $\ell$ as the horizontal integral length scale defined by
\begin{equation}
 \ell=2\pi\frac{\int_0^{+\infty}\frac{E(k_{\rm h})}{k_{\rm h}}{\rm d}k_{\rm h}}{\int_0^{+\infty}E(k_{\rm h}){\rm d}k_{\rm h}},
\end{equation}
where $E(k_{\rm h})$ is the horizontal spectrum of the turbulent kinetic energy averaged over the vertical direction, we find that the ratio $\ell/L_{\rm h}$ never exceeds 0.16 in our simulations.
This indicates that there are several large structures in the simulated domain.
For illustration, Fig.~\ref{fig_temperature} displays 2D snapshots of the temperature fluctuations and of the vertical velocity.

To properly resolve all scales of turbulence, we use a resolution of $128^2\times257$ and an aspect ratio $L_{\rm h}/L_{\rm v}=\pi/4$.
We have verified that in our simulation $k_{\rm max}\eta>1$,  where $k_{\rm max}$ is the largest wavevector present in the flow, $\eta=(\nu^3/\varepsilon)^{1/4}$ the Kolmogorov dissipation scale, and $\varepsilon$ the dissipation rate.
This criterion, which states that all scales down to the dissipation scale are resolved, has been found to be adapted for the study of turbulent transport in isotropic turbulence simulations \citep{Gotoh}.
In addition, simulations performed at a higher resolution have confirmed that our resolution is sufficient.

\section{Turbulent transport} \label{sec_transport}

The passive scalar is introduced once a statistical steady state is reached.
We then consider and compare two complementary approaches to determine its vertical turbulent transport: one Lagrangian, by following fluid particles; the other Eulerian, by solving an advection/diffusion equation for a concentration field.

Owing to the stationarity and the spatial homogeneity of the turbulence, a fluid particle encounters statistically homegeneous conditions as it moves with the flow
(as long as it keeps away from the upper and lower boundaries). The turbulence is thus homogeneous from a Lagrangian point of view.
This is an important property because it enables us to apply Taylor's turbulent transport theory \citep{Taylor}.
Accordingly, the transport is diffusive, and the vertical turbulent diffusion coefficient $D_{\rm t}$ reads as
\begin{equation} \label{eq_D_corr}
 D_{\rm t}=\int_0^{+\infty}\langle w(t)w(t+\tau)\rangle{\rm d}\tau=\langle w^2\rangle T_{\rm L},
\end{equation}
where $T_{\rm L}$ is the Lagrangian correlation time and $\langle\rangle$ the ensemble average over the particles.
If $z(t)$ denotes the vertical position of a particle, the vertical dispersion is then given by
\begin{equation} \label{eq_D_depl}
 \left\langle [z(t)-z(0)]^2\right\rangle=2D_{\rm t}t.
\end{equation}

We computed the turbulent diffusivity $D_{\rm t}$ using either Eq.~\eqref{eq_D_corr} or a linear regression of the mean square displacement $\left\langle [z(t)-z(0)]^2\right\rangle$.
In both cases the main source of error is the temporal fluctuations of the averaged turbulent quantities.
Then, depending on the starting point of the time average, this creates a significant dispersion (up to $20 \%$) in the values of $D_{\rm t}$.
We favor the turbulent diffusivity $D_{\rm t}$ obtained with Eq.~\eqref{eq_D_corr} because
in this case the dispersion is generally lower than using the linear regression of the mean square displacement.

In the Eulerian approach, the concentration field $c$ is governed by the equation
\begin{equation} \label{eq_conc}
 \frac{\partial c}{\partial t}+\vec u\cdot\vec\nabla c=D_{\rm m}\Delta c,
\end{equation}
where the molecular diffusivity $D_{\rm m}$ is such that $Pe_{\rm c}=L\Delta U/D_{\rm m}=10^4$.
To determine $D_{\rm t}$, we impose a mean concentration gradient ${\rm d}C/{\rm d}z$ and compute
\begin{equation} \label{eq_D_conc}
 D_{\rm t} = -\frac{\langle c'w\rangle}{{\rm d}C/{\rm d}z},
\end{equation}
where $c'$ refers to the concentration deviation from the mean profile $C(z)$.

We found that the relative difference between $D_{\rm t}$ computed from the Lagrangian approach (Eq.~\ref{eq_D_corr}) and the Eulerian one
does not exceed $15\%$. This difference is partly due to 
molecular diffusion, which is present in the Eulerian approach but not in the Lagrangian one. Nevertheless, this effect is limited
because $D_{\rm m}/D_{\rm t}$ remains always less than 5\%. 
Again, the difference is mostly due to temporal fluctuations of the turbulent quantities that generate dispersion in the time averages.
A way to reduce this dispersion is to average over a longer time. 
While it is not possible in the Lagrangian approach since the displacements of particles are 
limited by the upper and lower boundaries of the domain, a much longer integration time can be used in the Eulerian one.
We have thus used the Eulerian determination of the turbulent diffusion coefficient, keeping the error linked to the time average to a value lower than $5\%$.

\section{Results} \label{sec_results}

We present the results of five simulations corresponding to five different values of the turbulent P\'eclet number, $Pe_\ell=52, 0.90, 0.72, 0.34$, and $\ll 1$ (the last using the SPNA).
Table~\ref{tab_res} displays the turbulent diffusion coefficients, together with other relevant physical parameters.

\begin{table}
 \caption{Results of the different runs}
 \label{tab_res}
 \[
 \begin{array}{lccccc}
  \hline\hline
  Pe_\ell	&	Re_\ell	&	Ri_{\rm cr}	&	(RiPe_\ell)_{\rm cr}	&	Ri_{\rm cr}Pe_\ell	&	\beta=D_{\rm t}/(u\ell)\\ \hline
  52		&	260 	&	0.124		&	-   			        &	6.45		        &	0.104	\\
  0.90      &   150     &   1.10        &   -                       &   0.990               &   4.01\times10^{-2} \\
  0.72      &   240     &   1.07        &   -                       &   0.773               &   9.29\times10^{-2} \\
  0.34		&	340	    &	1.27		&	-   			        &	0.432		        &	0.138	\\
  \ll 1		&	335	    &	-		    &	0.426			        &	-   		        &	0.131	\\ \hline
 \end{array}
 \]
\end{table}

The most striking feature is that the full computation at $Pe_\ell=0.34$ and the SPNA computation give similar results.
Indeed, at $Pe_\ell=0.34$ the critical Richardson number multiplied by the turbulent P\'eclet number is very close to the critical ``Richardson-P\'eclet'' number of the SPNA computation.
This is strong evidence that at small P\'eclet number, the steady, stably stratified, sheared turbulence is characterized by a critical ``Richardson-P\'eclet'' number $(RiPe_\ell)_{\textrm{cr}}$ independent of the P\'eclet number.
This regime has not been reached in the simulations from $Pe_\ell=52$ to $Pe_\ell=0.7$ where $Ri_{\rm cr}Pe_\ell>(RiPe_\ell)_{\rm cr}$. 

Table~\ref{tab_res} further shows the parameter $\beta=D_{\rm t}/(u\ell)$, which in the regime where $RiPe_\ell=(RiPe_\ell)_{\textrm{cr}}$, verifies
\begin{equation} \label{eq_D_us}
 D_{\rm t} = \beta\kappa Ri^{-1}(RiPe_\ell)_{\textrm{cr}}.
\end{equation}
This parameter clearly has the same value for $Pe=0.34$ and $Pe\ll 1$.
Thus, in the small-P\'eclet-number regime, the turbulent diffusion coefficient can be written 
in the form of~\eqref{eq_D_us} with a constant $\beta$.
As a consequence, $D_{\rm t}$ is proportional to $\kappa Ri^{-1}$.

It is obvious that the final expression found for the vertical turbulent diffusion coefficient $D_{\rm t}$ (see Eq.~\ref{eq_D_us}) has the same form as Zahn's prescription 
given in Eq.~\eqref{eq_D_Zahn}, thus validating in the $Pe \le 0.34$ range the hypotheses made to derive it.
The present numerical simulations also provide a quantitative determination of the critical turbulent ``Richardson-P\'eclet'' number $(RiPe_\ell)_{\rm cr}=0.426$ and of the coefficient $\beta = D_{\rm t}/(u\ell) \simeq 0.131$.
This has to be compared with the order of magnitude estimates proposed by \cite{Zahn1992}, 
namely $Ri_{\rm c}\lesssim1/4$ and $\beta=1/3\simeq 0.333$. Nevertheless, it is the proportionality constant between $D_{\rm t}$ and $\kappa Ri^{-1}$, \emph{i.e.} the product of $\beta$ and $(RiPe_\ell)_{\rm cr}$, which is relevant in practice in stellar evolution codes.
We find a value of $5.58\times10^{-2}$ for this quantity, whereas Zahn's prescription leads to $8.33\times10^{-2}$, which is a relative difference of $33\%$.

\section{Conclusion and discussion} \label{sec_conclusion}

We thus conclude that the results of our DNS agree with the form of Zahn's prescription and provide
a first quantitative determination of $D_{\rm t}$ beyond Zahn's order of magnitude estimate.
In general, the main drawback of numerical simulations of turbulent flows is the unrealistically low value of the Reynolds number.
While it may be interesting to test the dependence of our results on the Reynolds number, we note
that estimates of effective turbulent Reynolds numbers associated with the radial diffusion of chemicals in stars lead
to surprisingly low values $Re_\ell \sim 100$ \citep{Michaud}, which are comparable to the values obtained in the present work. 

Other relevant issues can be addressed numerically.
In particular, the concentration field is considered here as a passive scalar so that the mean gradient has no effect on the flow, whereas in reality such a stratification would enhance the effect of temperature stratification.
This effect is considered differently by \cite{Maeder1996} and \cite{Talon}, and simulations might help in deciding between them.
Another important aspect of the models of rotationally-induced mixing is the magnitude of the horizontal turbulent transport driven by the horizontal differential rotation.
It controls the efficiency of the meridional circulation transport and the departure from shellular rotation \citep{Zahn1992}.
Again, the turbulent transport can be investigated with DNS similar to the present ones 
by varying the angle between the velocity and temperature gradients.

To study the turbulent transport in less idealized flow configurations would require a better understanding of the shear flows induced by the large-scale motions driven by rotation.

\begin{acknowledgements}
    This work was granted access to the HPC resources of CALMIP under the allocation 2012--P0507.
\end{acknowledgements}

\bibliographystyle{aa}
\bibliography{refs}

\end{document}